\documentclass[12pt]{article}

\usepackage{amssymb}
\usepackage{amsmath}
\usepackage{latexsym}
\usepackage{graphicx}

\catcode`@=11
\renewcommand{\section}{\@startsection{section}{1}{0pt}{\medskipamount}
{\medskipamount}{\large\bf}}
\numberwithin{equation}{section}
\catcode`@=12
\def\beq{\begin{eqnarray}}    %%%  begequation/eqnarray
\def\eeq{\end{eqnarray}}      %%%  endequation/eqnarray

\def\ln{\,\mbox{ln}\,}                  %%% log
\def\sTr{\,\mbox{sTr}\,}                %%% supertrace
\def\sDet{\,\mbox{sDet}\,}              %%% superdeterminant
\def\={\ =\ }

\begin{document}

\begin{titlepage}
\setcounter{page}{0}

\vskip 1.0cm

\begin{center}

{\LARGE\bf Field theories invariant under two-parametric supersymmetry}

\vspace{18mm}

{\Large S.R. Esipova$^{\dag}\footnote{E-mail:
esipova@tspu.edu.ru}$\;,
P.M. Lavrov$^{\dag\ddag}\footnote{E-mail:
lavrov@tspu.edu.ru}$
and
O.V. Radchenko$^{\dag\ddag}\footnote{E-mail: radchenko@tspu.edu.ru}$
}

\vspace{8mm}

\noindent  ${{}^\dag}
${\em
Tomsk State Pedagogical University,\\
Kievskaya St.\ 60, 634061 Tomsk, Russia}

\noindent  ${{}^\ddag}
${\em
National Research Tomsk State  University,\\
Lenin Av.\ 36, 634050 Tomsk, Russia}

\vspace{20mm}

\begin{abstract}
\noindent
We study field models for which a quantum action (i.e. the action appearing
in the generating functional of Green functions) is invariant under
supersymmetric transformations. We derive the Ward identity which is direct consequence
of this invariance. We consider a change of variables in functional integral connected
with supersymmetric transformations when its parameter is replaced by a nilpotent functional
of fields. Exact form of the corresponding Jacobian is found. We find restrictions on
generators of supersymmetric transformations when a consistent quantum description of
given field theories exists.
\end{abstract}

\end{center}

\vfill

\noindent {\sl Keywords:} supersymmetric invariance, field-dependent supersymmetric
transformation, nilpotency\\
\noindent {\sl PACS:} \ 04.60.Gw, \ 11.30.Pb

\end{titlepage}

%%%%%%%%%%%%%%%%%%%%%%%%%%%%%%%%%%%%%%%%%%%%%%%%%%%%%%%%%%%%%%

\section{Introduction and summary}

\noindent Field models with quantum action invariant under
supersymmetric transformations appear in several ways within modern
quantum field theory. The well-known example  is the Faddeev-Popov
action for Yang-Mills fields \cite{FP}, which  is invariant under
nilpotent supersymmetric transformations known as BRST
transformations \cite{brs,t}. Recent attempts \cite{S1,S2} to
formulate Yang-Mills fields in a form being  free of the Gribov
problem \cite{Gribov, Zwanziger1,Zwanziger2} give another examples
of actions invariant under some nilpotent supersymmetric
transformations. Superextension of sigma models \cite{CG} leads to
actions again invariant under supersymmetric transformations. Quite
recently a new realization of supersymmetry, called scalar
supersymmetry, has been proposed in \cite{J} when  one meets
supersymmetric invariant field models as well. The Curci-Ferrari
model of non-abelian massive vector fields \cite{CF} possesses
supersymmetric invariance connected with the modified BRST and
modified anti-BRST transformations. In contrast with the BRST
transformations these supersymmetric transformations are not
nilpotent. In turn it leads to serious consequences in physical
interpretation of the model \cite{L}.

In present paper from general point of view we study properties of
field theories for which an action appearing in the generating
functional of Green functions is invariant under supersymmetric
transformations. In  turn the supersymmetric transformations can  be
of three types. The first type is characterized as supersymmetric
transformations when there are no any restrictions on generators of
these transformations. We derive the Ward identity as a consequence
of the supersymmetric invariance and show that there is no a
possibility to present this identity in local form. The second type
consists of nilpotent supersymmetric transformations. Introducing
field-dependent nilpotent supersymmetric transformations we find an
exact form of the superdeterminant of the  change of variables in
the functional integral representing the generating functional of
Green functions. We prove that the action appearing after the change
of variables is not invariant under supersymmetric transformations
and find the source of this non-invariance. We consider the
non-invariance as inconsistency in the quantum presentation of given
theories. To lift the inconsistency we introduce the third type of
nilpotent supersymmetric transformations. In this case the
generators are subjected to an additional restriction.

We employ the condensed notation of DeWitt \cite{DeWitt}.
Derivatives with respect to fields are taken from the right.
Left derivatives with respect to fields are labeled
by a subscript~$l$. The Grassmann parity  of a quantity $X$ is denoted
as $\varepsilon(X)$. We use the notation $X_{,i}$
for right derivative of X with respect to $\phi^i$.
\\

\section{Supersymmetric invariant theories}

Our starting point is a theory of fields $\phi=\{\phi^i\}$ with
Grassmann parities $\varepsilon(\phi^i)=\varepsilon_i$. We assume a
non-degenerate action $S(\phi)$ of the theory so that the generating
functional of Green functions is given by the standard functional
integral
\beq
\label{ZJ}
Z(J)=\int {\cal
D}\phi\;\exp\Big\{\frac{i}{\hbar}\big[S(\phi)+J\phi\big]\Big\}\;.
\eeq We suppose invariance of $S(\phi)$
\beq
\label{Sinv}
S(\phi)=S(\varphi(\phi^{'}))=S(\phi^{'})
\eeq
under supersymmetric two-parametric transformations
\beq
%\nonumber
\label{sutr} \phi^i\quad \mapsto \quad \phi^i=\varphi^i(\phi^{'})\;,
\quad \varphi^i(\phi)=\phi^{i}+R^{ia}(\phi)\;\xi_a,\;\; a=1,2,
%\quad \varepsilon(R^i)=\varepsilon_i+1\;,
 \quad \xi_a\xi_b+\xi_b\xi_a=0\;,
\eeq
%Due to nilpotency of $\xi$ the relation (\ref{Sinv}) is equivalent to
so that
\beq
&&S_{,i}(\phi)R^{ia}(\phi)=0,
\qquad S_{,ij}(\phi)Z^{ji}(\phi)=0,\\
\nonumber
&&Z^{ij}(\phi)=\frac{1}{2}(-1)^{\varepsilon_j}
\varepsilon_{ab}\big(R^{jb}(\phi)R^{ia}(\phi)-
(-1)^{(\varepsilon_i+1)(\varepsilon_j+1)}R^{ib}(\phi)R^{ja}(\phi)\big)=0\;.
\eeq
In (\ref{sutr})
$\xi_a$ is  odd Grassmann parameters and $R^{ia}(\phi)$ are
generators of supersymmetric transformations having the Grassmann
parities opposite to fields $\phi^i$:
$\varepsilon(R^{ia})=\varepsilon_i+1$.

Consider now some consequence of the invariance on quantum level. To
this end we make the change of variables (\ref{sutr}) in the
functional integral (\ref{ZJ}). As a result we have
\beq Z(J)=\int
{\cal D}\phi\;\sDet M(\phi)\;
\exp\Big\{\frac{i}{\hbar}\big[S(\varphi(\phi))+J\varphi(\phi)\big]\Big\}
\eeq
where $\sDet M$ means the superdeterminant of supermatrix $M$
with matrix elements
\beq M^i_{\;j}(\phi)=\delta^i_{\;j}+
(-1)^{\varepsilon_i}\frac{\delta R^{ia}(\phi)}{\delta\phi^j}\xi_a\;,
\quad \varepsilon(M^i_{\;j})=\varepsilon_i+\varepsilon_j\;.
\eeq
In
general, for a theory under consideration this superdeterminant is
not equal to unity
\beq
\nonumber
&&\sDet M(\phi)=\exp\big\{\sTr\ln
M(\phi)\big\}= \exp\Big\{\frac{\delta
R^{ia}(\phi)}{\delta\phi^i}\xi_a+\Big\}=\\
&&= 1+\frac{\delta
R^{ia}(\phi)}{\delta\phi^i}\xi_a+=1+R^{ia}_{,i}(\phi)\xi_a+\;.
\eeq
It leads to
the following presentation of functional $Z(J)$ \beq \nonumber
Z(J)&=&\int {\cal D}\phi\;\Big(1+R^i_{,i}(\phi)\xi\Big)\;
\exp\Big\{\frac{i}{\hbar}\big[S(\phi)+J_i\phi^i+J_iR^i(\phi)\xi\big]\Big\}=\\
&=&\int {\cal D}\phi\;\Big(1+R^i_{,i}(\phi)\xi+\frac{i}{\hbar}J_iR^i(\phi)\xi\Big)\;
\exp\Big\{\frac{i}{\hbar}\big[S(\phi)+J\phi\big]\Big\}
\eeq
from which  the identity follows
\beq
\int {\cal D}\phi\;\Big(R^i_{,i}(\phi)+\frac{i}{\hbar}J_iR^i(\phi)\Big)\;
\exp\Big\{\frac{i}{\hbar}\big[S(\phi)+J\phi\big]\Big\}=0\;.
\eeq
With the help of usual manipulations this identity can be written in closed form
with respect to $Z(J)$
\beq
\label{WIZ}
\Big[J_iR^i\Big(\frac{\hbar}{i}\frac{\delta}{\delta J}\Big)-
i\hbar R^i_{,i}\Big(\frac{\hbar}{i}\frac{\delta}{\delta J}\Big)\Big]Z(J)=0\;.
\eeq
This identity is nothing but the Ward identity for generating functional of
Green functions. The existence of this identity is direct consequence of supersymmetric
invariance of $S(\phi)$. To simplify presentation of the Ward identity we
define the extended generating functional of Green functions by introducing
additional sources $K_i$ with Grassmann parities opposite to fields $\phi^i$,
$\varepsilon(K_i)=\varepsilon_i+1$
\beq
\label{ZJK}
Z(J,K)=\int {\cal D}\phi\;\exp\Big\{\frac{i}{\hbar}\big[S(\phi,K)+J\phi\big]\Big\}
\eeq
where
\beq
\label{SpK}
S(\phi,K)=S(\phi)+K_iR^i(\phi)\;,
\eeq
In general, the action $S(\phi,K)$ is not invariant under supersymmetric transformation
(\ref{sutr})
\beq
\label{SpKnon}
{\hat s}S(J,K)=K_i\;{\hat s}R^i(\phi)\neq 0\;,
\eeq
where the operator ${\hat s}$ of supersymmetric transformation was used.
Action of this operator on
arbitrary functional $X$ is given by the rule
\beq
{\hat s}X=\frac{\delta X}{\delta\phi^i}R^i\;.
\eeq

It is clear that there is the relation between functionals (\ref{ZJ}) and (\ref{ZJK})
\beq
Z(J,K)\big\vert_{K=0}=Z(J)\;.
\eeq

In terms of $Z(J,K)$ the Ward identity (\ref{WIZ}) reads
\beq
\label{WIZK}
J_i\frac{\delta Z(J,K)}{\delta K_i}=
i\hbar R^i_{,i}\Big(\frac{\hbar}{i}\frac{\delta}{\delta J}\Big)Z(J,K)\;.
\eeq
Note that the left side of the Ward identity (\ref{WIZK}) has the local form
in contrast with corresponding term in (\ref{WIZ}). In turn the right side of
(\ref{WIZK}) is a nonlocal.
\\

\section{Field-dependent supersymmetric transformations}

In this section we study more general type of supersymmetric transformations when
the parameter $\xi$ in (\ref{sutr}) is replaced by a field-dependent functional
$\xi(\phi)$
\beq
\label{sutrFD}
\varphi^i(\phi)=\phi^i+R^i(\phi)\xi(\phi)\;, \quad \xi^2(\phi)=0\;.
\eeq
We will referee to these transformations as field-dependent supersymmetric
transformations. Note that the action $S=S(\phi)$ remains invariant under
transformations (\ref{sutrFD}) due to nilpotency of $\xi(\phi)$
\beq
S(\phi)=S(\varphi(\phi^{'}))=S(\phi^{'})\;.
\eeq

Using the technique described in \cite{LL} it is not difficult to find the explicit
form of the superdeterminant of supermatrix
\beq
M^i_{\;j}(\phi)=\delta^i_{\;j}+R^i(\phi)\xi_{,j}(\phi)+
(-1)^{\varepsilon_i}R^i_{,j}(\phi)\xi(\phi)\;,
\eeq
corresponding to transformations (\ref{sutrFD}) with the result
\beq
\label{sDetFD}
\sDet M(\phi)=\big(1+{\hat s}\xi(\phi)\big)^{=1}
\Big[1+R^i_{,i}(\phi)\xi(\phi)-
\frac{\big({\hat s}^2\xi(\phi)\big)\xi(\phi)}{1+{\hat s}\xi(\phi)}\Big]\;.
\eeq
In (\ref{sDetFD}) we took into account that the action of the square operator ${\hat s}$
on an arbitrary functional $X$ is given by the relation
\beq
{\hat s}^2X=\frac{\delta X}{\delta\phi^i}\frac{\delta R^i}{\delta\phi^j}R^j=
X_{,i}R^i_{,j}R^j\;.
\eeq

In what follows we restrict ourselves to the case when the operator ${\hat s}$
is nilpotent, ${\hat s}^2=0$
\beq
\label{snil}
{\hat s}^2=0 \qquad \rightarrow \qquad \frac{\delta R^i}{\delta\phi^j}R^j=0\;.
\eeq
In particular, it means
\beq
{\hat s}R^i=0
\eeq
and we find that the action  $S(\phi, K)$ (\ref{SpK}), (\ref{SpKnon}) is invariant under
field-dependent supersymmetric transformations (\ref{sutrFD})
\beq
\label{SpKinv}
S(\phi, K)_{,i}\;R^i(\phi)=0\;.
\eeq
The invariance of $S(\phi, K)$ can be expressed in an unique form
\beq
\label{SpKinv1}
\frac{\delta S(\phi,K)}{\delta\phi^i}\;\frac{\delta S(\phi,K)}{\delta K_i}=0\;.
\eeq
The equation (\ref{SpKinv1}) is nothing but the Zinn-Justin equation appearing for
the first time in quantization of non-abelian gauge fields \cite{Z-J}.

Performing the change of variables in form of field-dependent supersymmetric
transformations (\ref{sutrFD}), (\ref{snil}) and using (\ref{sDetFD}) we have
\beq
\sDet M(\phi)=\exp\big\{R^i_{,i}(\phi)\xi(\phi)-
\ln\big(1+{\hat s}\xi(\phi)\big)\big\}=
\big(1+{\hat s}\xi(\phi)\big)^{-1}\big[1+R^i_{,i}(\phi)\xi(\phi)\big]
\eeq
and arrive at the following presentation of generating functional $Z(J,K)$
\beq
\nonumber
\label{ZJKtr}
Z(J,K)&=&\int {\cal D}\phi\;
\exp\Big\{\frac{i}{\hbar}\big[S(\phi,K)+J\big(\phi+R(\phi)\xi(\phi)\big)-\\
&&\qquad\qquad\qquad -i\hbar R^i_{,i}(\phi)\xi(\phi)+
i\hbar\ln(1+{\hat s}\xi(\phi))\big]\Big\}\;.
\eeq
We can rewrite the presentation (\ref{ZJKtr}) in the form
\beq
Z(J,K)=Z(J,K)+I(J,K)
\eeq
where
\beq
\nonumber
I(J,K)&=&\int {\cal D}\phi\;\big(1+{\hat s}\xi(\phi)\big)^{-1}
\Big[{\hat s}\xi(\phi)-R^i_{,i}(\phi)\xi(\phi)-
\frac{i}{\hbar}J_iR^i(\phi)\xi(\phi)\Big]\times\\
&&\qquad\qquad\;\times\exp\Big\{\frac{i}{\hbar}\big[S(\phi,K)+J\phi\big]\Big\}\;.
\eeq
The functional $I(J,K)$ should be zero. Let us prove this property. To this end it
is useful to introduce the functional $\Lambda(\phi)$
\beq
\Lambda(\phi)=\xi(\phi)\big(1+{\hat s}\xi(\phi)\big)^{-1}\;,
\eeq
so that
\beq
{\hat s}\Lambda(\phi)=\frac{{\hat s}\xi(\phi)}{1+{\hat s}\xi(\phi)}\;.
\eeq
Then we have
\beq
\nonumber
I(J,K)&=&\int {\cal D}\phi\;
\Big[\frac{\delta\Lambda(\phi)}{\delta\phi^i}R^i(\phi)+\Lambda(\phi)R^i_{,i}(\phi)
+\frac{i}{\hbar}\Lambda(\phi)J_iR^i(\phi)\Big]\times\\
\nonumber
&&\qquad\qquad\times\;\exp\Big\{\frac{i}{\hbar}\big[S(\phi,K)+J\phi\big]\Big\}=\\
&=&\int {\cal D}\phi\;\frac{\delta}{\delta\phi^i}
\Big[\Lambda(\phi)R^i(\phi)
\exp\Big\{\frac{i}{\hbar}\big[S(\phi,K)+J\phi\big]\Big\}\Big]=0\;
\eeq
where the invariance of $S(\phi, K)$ (\ref{SpKinv}) was used.

From (\ref{ZJKtr}) it follows that a theory with the action $S(\phi,K)$ invariant under
supersymmetric transformation (\ref{sutr}) or (\ref{sutrFD}) admits formulation in term
of action $S_{\xi}(\phi,K)$
\beq
\label{Sxi}
S_{\xi}(\phi,K)=S(\phi, K)+i\hbar\ln(1+{\hat s}\xi(\phi))-
i\hbar R^i_{,i}(\phi)\xi(\phi)
\eeq
In its turn, in general, the action $S_{\xi}(\phi,K)$ is not invariant under
supersymmetric transformations (\ref{sutr}) or (\ref{sutrFD}) due to the third term in
rhs (\ref{Sxi})
\beq
{\hat s}S_{\xi}(\phi, K)=-i\hbar\;{\hat s}\big(R^i_{,i}(\phi)\xi(\phi)\big)\neq 0\;.
\eeq
We consider this as an indication of the inconsistency in formulation of the model being
invariant under supersymmetric  transformations. Indeed, it seems strange that a theory
with the action invariant under supersymmetric transformations is equivalently presented
in the form when this symmetry looks like broken. This inconsistency can be deleted if
the additional requirement is fulfilled, $R^i_{,i}(\phi)=0$.
\\

\section{Special supersymmetric theories}

We will refer {\it special  supersymmetric theories} for such theories which are
invariant under nilpotent supersymmetric transformations when the generators $R^i(\phi)$
are subjected to the restriction\footnote[1]{In terms of paper
\cite{LSh} this restriction
means that a modular class of a given gauge systems vanishes.}
\beq
\label{Rad}
R^i_{,i}(\phi)=0\;.
\eeq
Note that the generators of BRST transformations in Yang-Mills theories satisfy
this relation. One can easily check that for generators of nilpotent supersymmetric
transformations appearing in models introducing in papers \cite{S1,S2,CG,J} the condition
(\ref{Rad}) is valid as well. In case of special supersymmetric theories
the superdeterminant of field-dependent supersymmetric transformations reads
\beq
\sDet M = \frac{1}{1+{\hat s}\xi}\;
\eeq
and the action (\ref{Sxi}) reduces to
\beq
\label{Sxir}
S_{\xi}(\phi,K)=S(\phi,K)+i\hbar\ln(1+{\hat s}\xi(\phi))\;.
\eeq
Using the nilpotency of ${\hat s}$ we can present the action (\ref{Sxir})
as the modification of initial action $S(\phi)$ by ${\hat s}$-exact term
\beq
\label{Sxir1}
S_{\xi}(\phi,K)=S(\phi,K)+{\hat s}F(\phi)=S(\phi)+{\hat s}\big(K\phi+F(\phi)\big)\;,
\eeq
where
\beq
F=\xi\Big[1-\frac{1}{2}({\hat s}\xi)+\frac{1}{3}({\hat s}\xi)^2-\cdots\Big]=
\xi \;({\hat s}\xi)^{-1}\;\ln (1+{\hat s}\xi)\;
\eeq
is a regular function. The presentation (\ref{Sxir1}) can be very useful in theories
with supersymmetric invariant action. In particular, it was shown \cite{LL} that
in case of Yang-Mills theories the result of change of
variables in vacuum functional with the help of field-dependent BRST transformations
can be presented in the form likes (\ref{Sxir1}) and interpreted as a modification of
gauge condition. This made it possible to prove
the independence of the effective action in Yang-Mills theories on
the finite increment of gauge on-shell
and suggest  the formulation of the Gribov-Zwanziger theory
\cite{Gribov,Zwanziger1,Zwanziger2} free from the problem of gauge dependence
(for details, see \cite{LLR}) of the effective action on-shell \cite{LL2}.

From(\ref{Sxir1}) it is clear invariance of $S_{\xi}(\phi, K)$ under supersymmetric
transformations
\beq
 {\hat s}S_{\xi}(\phi,K)=0\;.
\eeq
This invariance can be expressed in the form of Zinn-Justin equation
\beq
\label{Sinvxi}
\frac{\delta S_{\xi}}{\delta\phi^i}\frac{\delta S_{\xi}}{\delta K^i}=0\;.
\eeq
As a consequence of the equation (\ref{Sinvxi}) the generating functional
$Z_{\xi}(J,K)$ constructed with the help of action $S_{\xi}(\phi, K)$ satisfies
the Ward identity
\beq
\label{WIZxi}
J_i\frac{\delta Z_{\xi}(J,K)}{\delta K_i}=0
\eeq
as the functional $Z(J,K)$. One can rewrite the Ward identity (\ref{WIZxi})
in term of the generating functional of connected Green functions $W_{\xi}(J,K)$
\beq
Z_{\xi}(J,K)=\exp\Big\{\frac{i}{\hbar}W_{\xi}(J,K)\Big\}
\eeq
as
\beq
J_i\frac{\delta W_{\xi}(J,K)}{\delta K_i}=0\;.
\eeq
Making use the Legendre transformation
\beq
\phi^i=\frac{\delta W_{\xi}(J,K)}{\delta J_i}
\eeq
and introducing the generating functional of vertex functions $\Gamma_{\xi}(\phi,K)$
\beq
\Gamma_{\xi}(\phi,K)=W_{\xi}(J,K)-J_i\phi^i\;, \qquad
\frac{\delta\Gamma_{\xi}}{\delta K_i}=\frac{\delta W_{\xi}}{\delta K_i}\;,\qquad
\frac{\delta\Gamma_{\xi}}{\delta\phi^i}=-J_i\;,
\eeq
the Ward identity for $\Gamma_{\xi}=\Gamma_{\xi}(\phi,K)$
\beq
\label{WIGxi}
\frac{\delta\Gamma_{\xi}}{\delta \phi^i}\frac{\delta\Gamma_{\xi}}{\delta K_i}=0
\eeq
has the form of the Zinn-Justin equation and repeats on quantum level the invariance
of a given theory under supersymmetric transformations. It is clear that all relations
(\ref{WIZxi})-(\ref{WIGxi}) are valid for initial theory ($\xi=0$).
\\

\section*{Acknowledgments}
\noindent   The work  is partially supported  by the Ministry of
Education and Science of Russian Federation, grant TSPU-122.
The work of PML and OVR is partially supported by the RFBR grant
12-02-00121 and the Presidential grant 88.2014.02 for LRSS.

%\newpage

\end{document}